\documentclass[12pt]{article}

\usepackage{amsmath,amssymb,amsfonts,amsbsy}
\usepackage{cite}
\usepackage{graphicx}
\usepackage{wrapfig}

\begin{document}
\addcontentsline{toc}{subsection}{{Title of the article}\\
{\it B.B. Author-Speaker}}

\setcounter{section}{0}
\setcounter{subsection}{0}
\setcounter{equation}{0}
\setcounter{figure}{0}
\setcounter{footnote}{0}
\setcounter{table}{0}

\begin{center}
\textbf{MICROSCOPIC STERN-GERLACH EFFECT AND THOMAS SPIN PRECESSION
 AS AN ORIGIN OF THE SSA}

\vspace{5mm}

 \underline{V.V.~Abramov}$^{\,1}$ 
\vspace{5mm}

\begin{small}
  (1) \emph{Institute for High Energy Physics, Protvino, Russia} \\
  $\dag$ \emph{E-mail: Victor.Abramov@ihep.ru}
\end{small}
\end{center}

\vspace{0.0mm} 

\begin{abstract}
The single-spin asymmetry and hadron polarization data are analyzed 
in the framework of a phenomenological effective-color-field model. 
Global analysis of the single-spin effects in hadron production is 
performed for h+h, h+A, A+A and  lepton+N interactions. The model 
explains the dependence of the data on $x_{F}$, $p_{T}$, collision 
energy $\sqrt{s}$ and atomic weights $A_{1}$ and $A_{2}$ of colliding 
nuclei. The predictions are given for not yet explored kinematical regions.
\end{abstract}

\vspace{7.2mm} 

In this report we discuss a semi-classical mechanism for the 
single-spin phenomena in inclusive reaction $A + B \rightarrow C + X$. 
The major assumptions of the model are listed below.

1)	An effective color field (ECF) is a superposition of the QCD string
 fields, created by spectator quarks and antiquarks after the initial color
 exchange.

2) The constituent quark $Q$ of the detected hadron $C$ interacts with the 
nonuniform chromomagnetic field via its chromomagnetic moment 
$\mu^{a}_{Q} = sg^{a}_{Q}g_{S}/2M_{Q}$ and with the chromoelectric 
field via its color charge $g_{S}$.

3) The microscopic Stern-Gerlach effect in chromomagnetic field 
$\rm\bf{B^{a}}$ and Thomas spin precession in chromoelectric field 
$\rm\bf{E^{a}}$ lead to the large observed SSA. The ECF is considered 
as an external with respect to the quark $Q$ of the observed hadron.

4) The quark spin precession in the ECF (chromomagnetic and chromoelectric) 
is an additional phenomenon, which leads to the specific SSA dependence 
(oscillation) as a function of kinematical variables ($x_{F}$, $p_{T}$ or
 scaling variable $x_{A}=(x_{R}+x_{F})/2)$.

The longitudinal field $\rm\bf{E^{a}}$  and the circular  field 
$\rm\bf{B^{a}}$ of the ECF are written as
\begin{equation}\label{eq:Field-B}
 E^{(3)}_{Z} = -2\alpha_{s}\nu_{A} /\rho^{2}exp(-r^{2}/\rho^{2}) \,,\quad  
 B^{(2)}_{\varphi} = -2\alpha_{s}\nu_{A} r/\rho^{3}exp(-r^{2}/\rho^{2}) \,, 
\end{equation}
where $r$ is the distance from the string axis, $\nu_{A}$ is the number of 
quarks at the end of the string, 
$\rho = 1.25 R_{c}$,   $R_{c}$  - confinement radius, $\alpha_{s}\approx 1$
 - running coupling constant\cite{Migdal,YF09}.

The Stern-Gerlach type forces act on a quark moving inside the ECF (flux tube):
\begin{equation}\label{eq:force}
    f_{x} =  \mu^{a}_{x}\partial B_{x}^{a}/\partial x  +
   \mu^{a}_{y}\partial B_{y}^{a}/\partial x  \,,\quad
    f_{y} =  \mu^{a}_{x}\partial B_{x}^{a}/\partial y  +
   \mu^{a}_{y}\partial B_{y}^{a}/\partial y \,. 
\end{equation}

The quark $Q$ of the observed hadron $C$, which gets $p_{T}$ kick of
 Stern-Gerlach forces and undergo a spin precession in the ECF is called
 a ``probe'' and it ``measures'' the fields  $\rm\bf{B^{a}}$  and 
$\rm\bf{E^{a}}$.
The ECF is created by spectator quarks and antiquarks and obeys quark 
counting rules. 
Spectators are all quarks which are not constituents of the observed
 hadron $C$.	In the case of  $p^{\uparrow} + p \rightarrow \pi^{+} + X$ 
reaction (see Fig. 1) polarized probe  $u$ quark from $\pi^{+}$ feels 
the field, created by the spectator quarks with weight 
$\lambda  = -|\Psi_{qq'}(0)|^{2}/|\Psi_{q\bar{q}'}(0)|^{2} \approx -1/8$, 
by antiquarks with weight 1, and by target quarks with weight $-\tau\lambda$,
 respectively. Spectator quarks from the target $B$ have an additional
 negative factor $-\tau= -0.0562\pm 0.030$, since these quarks are moving 
in opposite direction in cm reference frame.  The value of color factor
 $\lambda = -0.1321\pm 0.0012$, obtained in a global fit of 68 inclusive 
reactions, is close to the expected one, which is a strong argument in
 favor of the ECF model\cite{YF09}.

Another important phenomenon is quark spin precession in the ECF.
We assume that the spin precession is described by the 
Bargman-Michel-Telegdi eqs. (3)-(4)\cite{BMT}: 
\begin{equation}
d{\rm\bf  \xi } /d t = 
 a[{\rm\bf \xi B^{a} } ] 
 +d [ {\rm\bf \xi [ E^{a}  v ] }],
\label{eq:dSdT}
\end{equation}
\begin{equation}\label{eq:a}
  a = g_{S}(g^{a}_{Q} - 2 + 2M_{Q}/E_{Q})/2M_{Q} \,,\quad
d =  g_{S} [ g^{a}_{Q} - 2E_{Q}/(E_{Q}+M_{Q}) ]/2M_{Q} \,.
\end{equation}
  \vspace*{+2mm} 
\begin{wrapfigure}[12]{R}{60mm}
  \centering 
  \vspace*{-8mm} 
  \includegraphics[width=60mm]{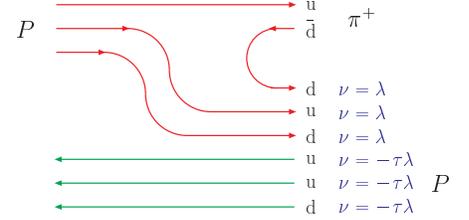}
  \vspace*{+2mm} 
  \caption{The quark flow diagram for the reaction 
$p^{\uparrow} + p \rightarrow \pi^{+} + X$. The weight $\nu$ of each
 spectator quark contribution to the ECF is indicated.} 
  \label{abramov_fig1}
\end{wrapfigure}
The precession frequency depends on the color charge $g_{S}$, the quark
 mass $M_{Q}$ and its energy $E_{Q}$, and on the color $g^{a}_{Q}$-factor.
The value  $\Delta\mu^{a}  = (g^{a}_{Q}-2)/2$ is called a color anomalous 
magnetic moment and it is large and negative in the instanton model.
 Spontaneous chiral symmetry breaking leads to an additional dynamical mass
 $\Delta M_{Q}(q)$  and  $\Delta\mu^{a}(q)$, both depend on momentum 
transfer $q$\cite{Diakonov}. 
Kochelev predicted $\Delta\mu^{a}(0) = -0.2$\cite{Kochelev} and Diakonov
 predicted $\Delta\mu^{a}(0) =-0.744$\cite{Diakonov}.
The global data analysis results are closer to the Diakonov's predictions.

Due to the microscopic Stern-Gerlach effect the probe quark $Q$ gets an 
additional spin-dependent transverse momentum $\delta p_{x}$, which causes 
an azimuthal asymmetry or observed hadron polarization:
\begin{equation}
\delta p_{x} =
{ { {g^{a}_{Q}   \xi^{0}_{y}}         
 \over 
  {2\rho (g^{a}-2 + 2M_{Q}/E_{Q} ) } }  [
{ {1 - \cos(\phi_{A}) } \over {\phi_{A}}  }} +\epsilon\phi_{A} ],
\label{eq:dpt2}
\end{equation}
where $\phi_{A}=\omega_{A}x_{A}$ is a quark spin precession angle in 
the fragmentation region of the beam particle $A$, and a  "frequency" 
 of $A_{N}$ oscillation as a function of $x_{A}$ is 
\begin{equation}\label{eq:omega0A}
\omega_{A} =
 {  { g_{S} \alpha_{s} \nu_{A} S_{0} (g^{a}_{Q}-2 + 2M_{Q}/E_{Q}) }
 \over { M_{Q}\rho^{2}c}  }   \,.
\end{equation}
The  length $S_{0}$ of the ECF is $0.6 \pm 0.2$ fm. The constituent quark
 masses $M_{Q}$ and $\Delta\mu^{a}$ values are given in \cite{YF09}. 
The parameter $\epsilon = -0.00419 \pm 0.00022$
is small due to subtraction of the Thomas precession term from 
$\epsilon = 1/2$ for chromomagnetic contribution to the $\delta p_{x}$.

Let us consider in more detail the Thomas precession effect in the ECF.
Due to the Thomas precession an additional term 
$U= {\rm\bf s\cdot \omega_{T}}$ appears in the effective Hamiltonian.
The Thomas frequency ${\rm\bf \omega_{T} \approx [F v]/}M_{Q} $ depends 
on the force $\rm\bf F$, the quark velocity $\rm\bf v$ and its mass $M_{Q}$.
The quark polarization due to the Thomas precession,
${\rm\bf \delta P_{N}= -\omega_{T} /}\Delta E$, is directed opposite 
to the frequency vector ${\rm\bf \omega_{T}}$ direction since  
$\Delta E$ is positive\cite{DM}.

The sign and magnitude of the force ${\rm\bf F}= g_{S}{\rm\bf E^{a}}$ 
is determined by the quark-counting rules for the ECF. For example, 
$F_{Z} \approx -2g_{S}\alpha_{S}[1+\lambda -3\tau\lambda ]/\rho^{2} < 0$
for the $pp  \rightarrow \Lambda^{\uparrow} + X$ reaction at 
$\sqrt{s}<70$ GeV. 
In case of the reactions $pp  \rightarrow \Xi^{0 \uparrow} + X$ and 
$p^{ \uparrow} + p \rightarrow \pi^{+} + X$ the factor in square brakets 
is $[2+2\lambda -3\tau\lambda ] <0$ and $[3\lambda -3\tau\lambda]>0$, 
respectively. 

 Recombination potential is more attractive for negative 
$U= {\rm\bf s\cdot \omega_{T}}$. As a result, for the
 $pp  \rightarrow \Lambda^{\uparrow} + X$ reaction  the additional 
 Thomas precession contribution to $P_{N}$ is positive and opposite 
in sign to the dominating chromomagnetic term, which gives 
$P_{N} <0$, and to the DeGrand-Miettinen model predicted $P_{N}<0$ \cite{DM}.
 The additional transverse momentum $\delta p_{x}$ is related 
to the analyzing power or polarization by the relation 
 $A_{N}=-D \delta p_{x}$,
where $D=5.68 \pm 0.13$ GeV$^{-1}$ is an effective slope of
 the invariant cross section.
In the Ryskin model $\delta p_{x} \approx  0.1$ GeV/c  is a constant 
\cite{Rysk88}.   
In the ECF model we have dynamical origin of $A_{N}$ or $P_{N}$ dependence
 on the kinematical variables
 ($p_{T}$, $x_{A}$, $x_{B}= (x_{R}-x_{F})/2$, $x_{F}$)  and on the number of
 (anti)quarks in hadrons $A$, $B$ and $C$, and also on quark 
$g^{a}_{Q}$-factor and its mass $M_{Q}$. This dependence is due to 
the microscopic Stern-Gerlach effect and quark spin precession in 
the ECF \cite{YF09}.
%
%
\begin{figure}[hb!]
  \centering
  \begin{tabular}{cc}
    \includegraphics[width=60mm]{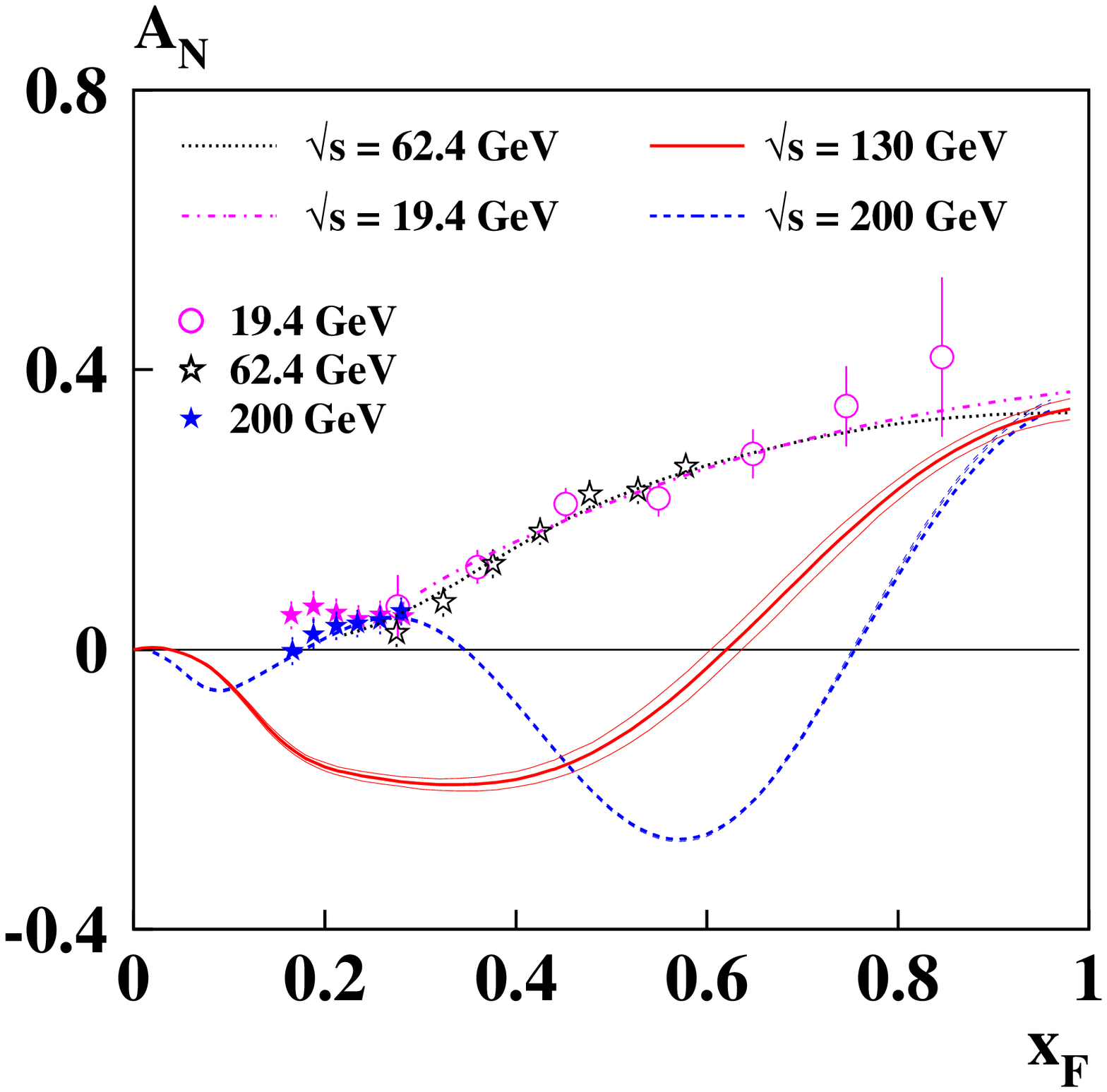} &
     \includegraphics[width=60mm]{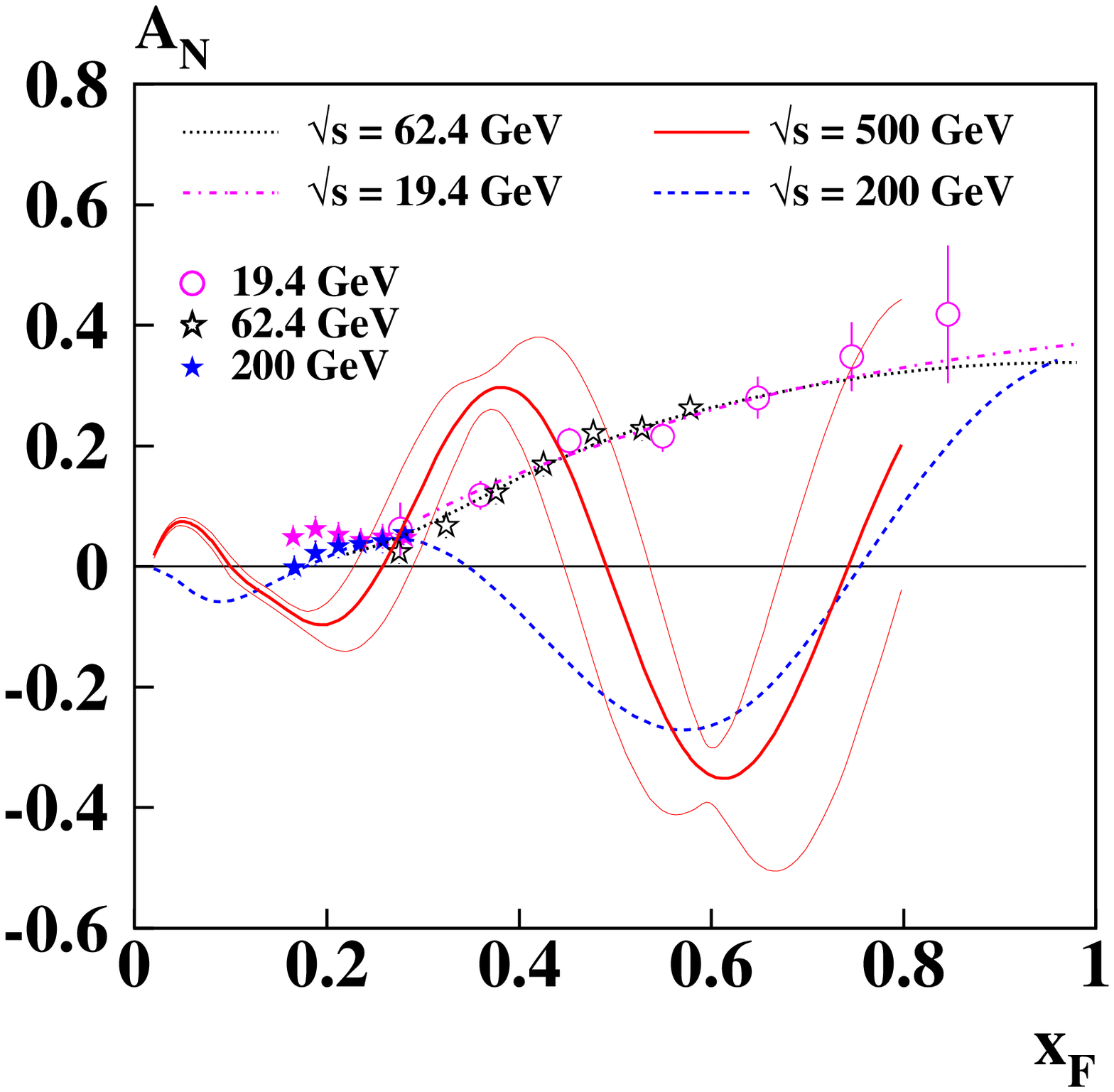} \\
    \textbf{(a)} & \textbf{(b)} 
  \end{tabular}
  \caption{%
     The dependence $A_{N}(x_{F})$ for
 $p^{ \uparrow} + p \rightarrow \pi^{+} + X$ reaction. 
Predictions are for \textbf{(a)} $\sqrt{s}=$ 130 GeV and
 \textbf{(b)} $\sqrt{s}=$ 500 GeV, respectively. The cm 
 production angle is $4.1^{o}$.
  }
  \label{abramov_fig2}
\end{figure}
%

The ECF increases dramatically at energy $\sqrt{s} > 70$ GeV  or in
 collisions of nuclei.
 In the case of nuclei collisions the effective number of quarks in 
a projectile nuclei, which contributes to the ECF, is equal to its 
number in a tube with transverse radius limited by the confinement.
The new quark contribution to the ECF depends on kinematical variables: 
\begin{equation}\label{eq:fN}
 f_{N} = n_{q}\exp(-W/\sqrt{s})(1-x_{N})^{n}  \,,\quad
x_{N} = [ (p_{T}/p_{N})^{2} + x^{2}_{F} ]^{1/2}  \,,
\end{equation}
where $W \approx 238(A_{1}A_{2})^{-1/6}$ GeV, 
$n \approx 0.91(A_{1}A_{2})^{1/6}$,  $p_{N}\approx 28$ GeV/c and
$n_{q} \approx 4.52$ \cite{YF09}.

  \vspace*{+2mm} 
\begin{wrapfigure}[12]{R}{50mm}
  \centering 
  \vspace*{-8mm} 
  \includegraphics[width=55mm,height=50mm]{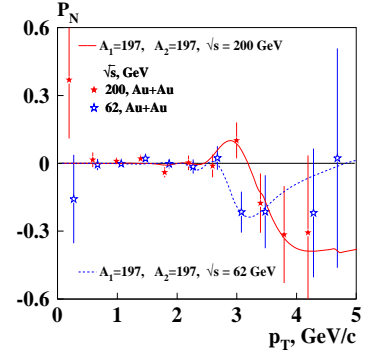}
  \caption{The dependence $P_{N}(p_{T})$ of $\Lambda$-hyperon  polarization 
in Au+Au collisions.} 
  \label{abramov_fig3}
\end{wrapfigure}
%

Due to the dependence of the ECF on $\sqrt{s}$  and atomic weights $A_{1}$,
 $A_{2}$ of colliding particles we expect very unusual behavior of $A_{N}$ 
and $P_{N}$ as a function of kinematical variables. The data and model 
predictions of $A_{N}$ for the $\pi^{+}$  production in pp-collisions are
 shown in Fig.~2a as a function of $x_{F}$. The data are from the E704 
\cite{E704} and BRAHMS \cite{BRAHMS} experiments for cm energies 
19, 62 and 200 GeV. The model predictions describe the data. The dashed curve
 is for 200 GeV and the solid curve is for 130 GeV. We expect a negative 
$A_{N}$ for 200 GeV and $x_{F}$ around 0.6 and also for 130 GeV and $x_{F}$ in
 the range from 0.1 to 0.6. The negative $A_{N}$ values are due to the 
$u$-quark spin precession in a strong ECF.  Even more unusual oscillating 
behavior of $A_{N}$ is expected for 500 GeV (see Fig.~2b).

In Fig.~3 the global polarization $P_{N}$ of $\Lambda$ hyperon in Au+Au
 collisions is shown as a function of $p_{T}$. The data are from STAR 
experiment at 62 and 200 GeV \cite{STAR}. The ECF model predictions 
reproduce the data. Oscillating behavior of $P_{N}$ is due to the 
s-quark spin precession in very strong color fields.
Similar oscillating behavior of $P_{N}$, as a function of  pseudorapidity,
 is expected for low energies, 7 and 9 GeV in cm. The predictions are shown
 in Fig.~4
for S+S, Cu+Cu,  and Au+Au collisions, respectively.

Conclusion:
             a semi-classical mechanism is proposed for single-spin phenomena.
 The effective color field of QCD strings, created by spectator quarks and
 antiquarks is described by the quark-counting rules. The microscopic 
Stern-Gerlach effect in the chromomagnetic field and the Thomas spin
 precession in the chromoelectric field lead to the SSA. 
 The energy and atomic weight dependence of the effective color fields, 
combined with the quark spin precession phenomenon, lead to the oscillating
 behavior of $A_{N}$ and $P_{N}$. 	The model predictions for different
 reactions and energies can be checked at the existing accelerators.
\begin{figure}[b!]
  \centering
  \begin{tabular}{ccc}
  \hspace*{-6mm} 
    \includegraphics[width=55mm]{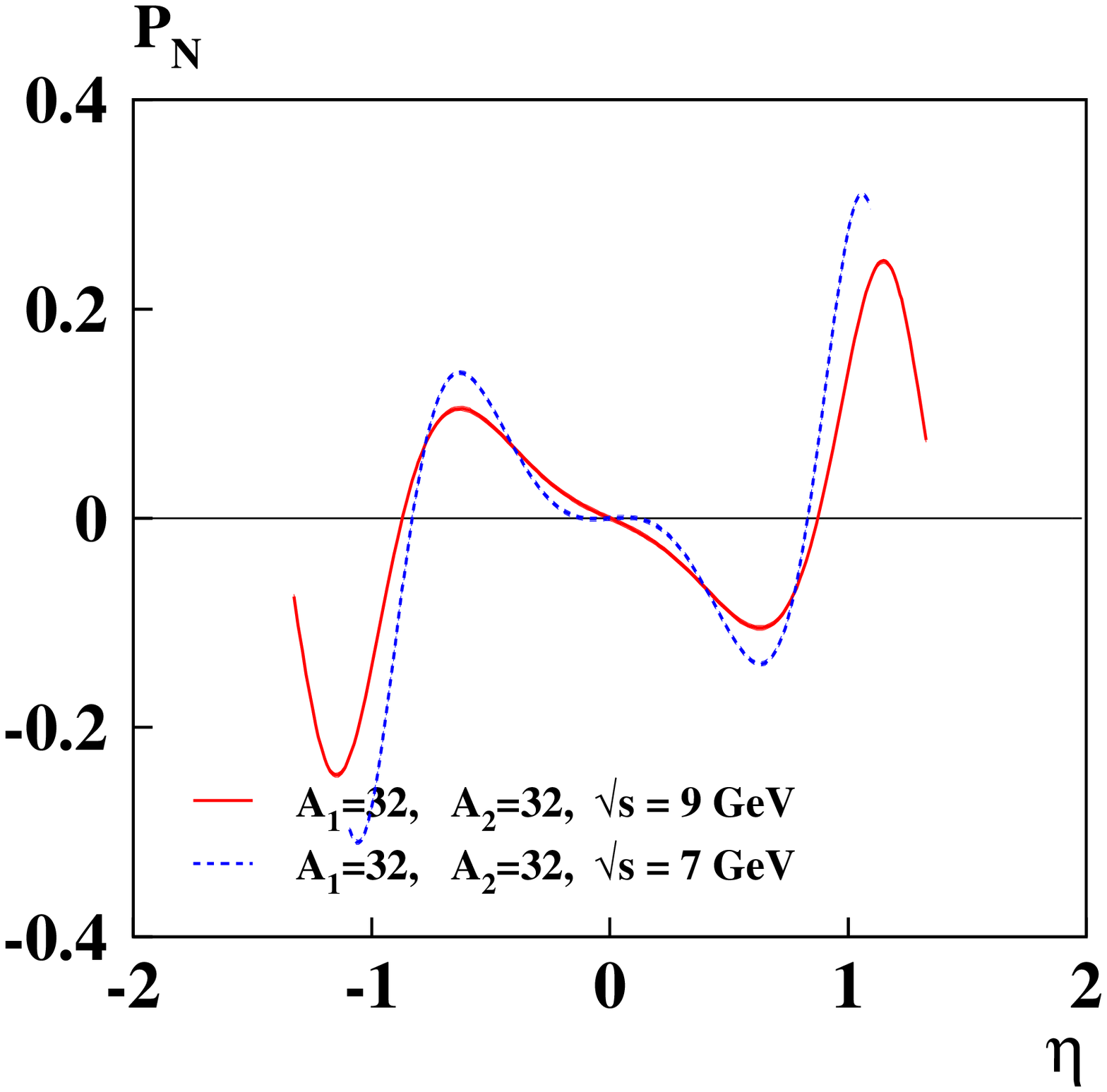} &
  \hspace*{-6mm} 
    \includegraphics[width=55mm]{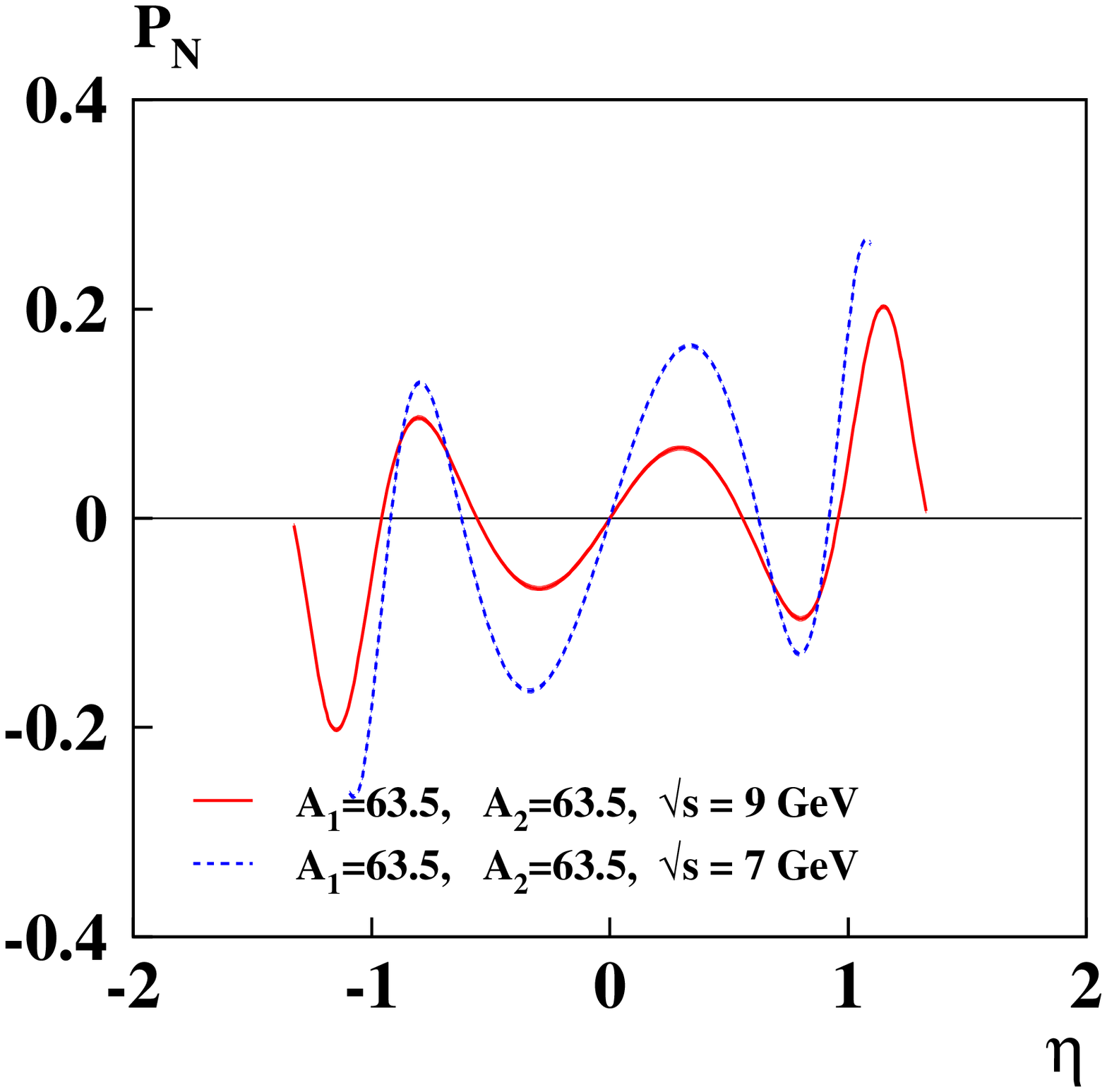} &
  \hspace*{-6mm} 
    \includegraphics[width=55mm]{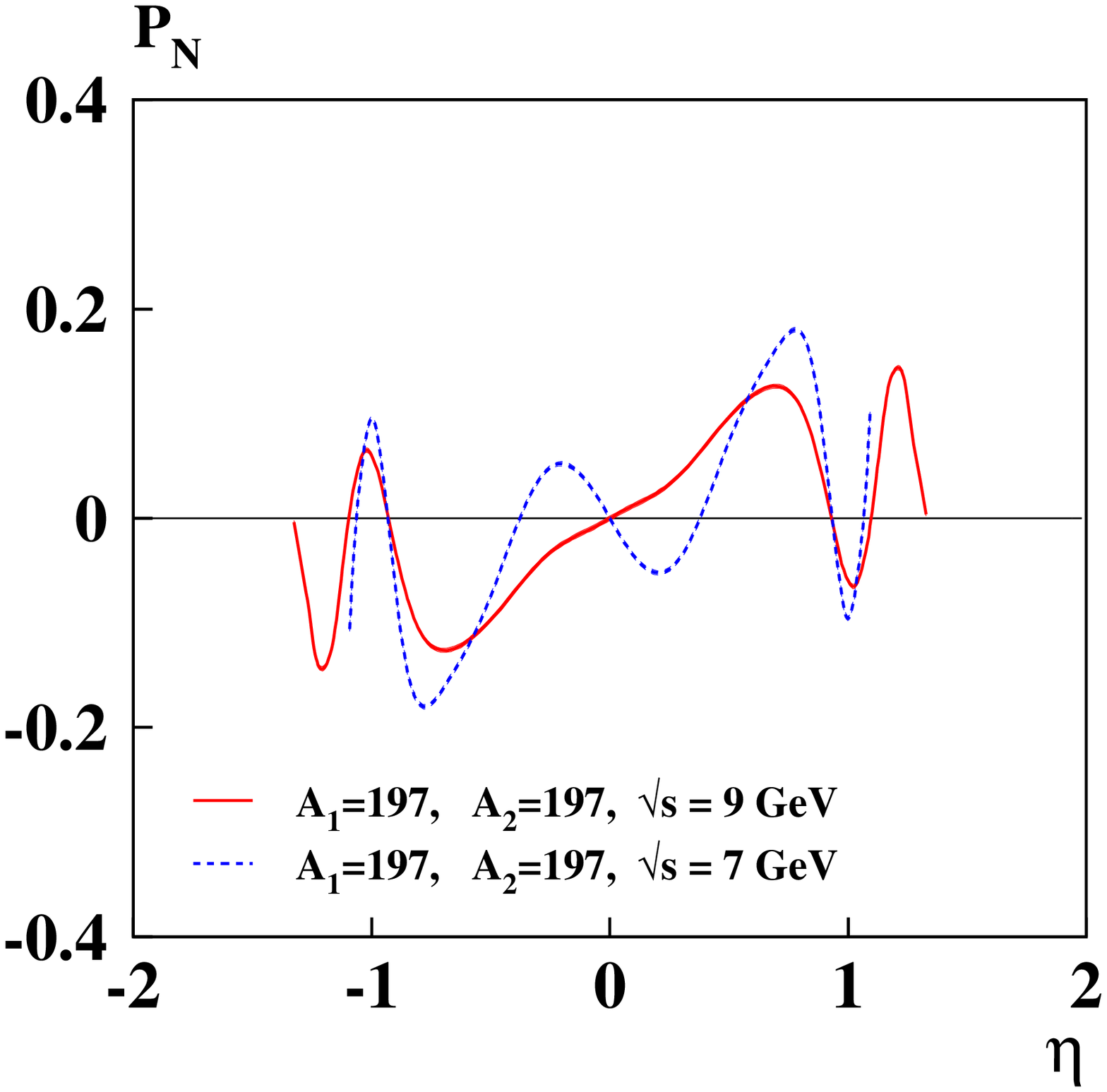} \\
    \textbf{(a)} & \textbf{(b)} & \textbf{(c)}
  \end{tabular}
  \caption{%
    The transverse polarization of $\Lambda$ vs $\eta=-ln(\tan\theta_{CM}/2)$ 
at $p_{T}=2.35$ GeV/c in  \textbf{(a)} S+S,  \textbf{(b)} Cu+Cu  and 
 \textbf{(c)} Au+Au  collisions. Solid line corresponds to $\sqrt{s}=9$ GeV 
and dashed line to $\sqrt{s}=7$ GeV, respectively.
  }
  \label{abramov_fig4}
\end{figure}
%
%

\end{document}